\documentclass{article}
\usepackage{spconf,amsmath,graphicx}
\usepackage{subcaption}
\usepackage{amsthm}
\usepackage{amsfonts}
\usepackage{enumitem}
\usepackage{amssymb}
\usepackage{xcolor}
\usepackage{url}
\usepackage{graphicx}
\usepackage{subcaption}
\usepackage[ruled,vlined]{algorithm2e}
\usepackage{xcolor}
\usepackage[normalem]{ulem}
\usepackage{amsfonts}
\usepackage{amsmath}
\usepackage{amssymb}
\usepackage{xcolor}
\usepackage{amsthm}

\newtheorem{assumption}{Assumption}[section]
\newtheorem{theorem}[assumption]{Theorem}

\DeclareMathOperator*{\argmin}{arg\,min}

\usepackage{hyperref}

\newcommand{\E}[0]{\mathbb{E}}

\newcommand{\Cov}[0]{\operatorname{Cov}}
\newcommand{\R}[0]{\mathbb{R}}

\usepackage{setspace}
\renewenvironment{thebibliography}[1]{%
	\begin{oldthebibliography}{#1}%
		\setlength{\itemsep}{-0.35ex}%
	}%
	{%
	\end{oldthebibliography}%
}

\title{Generalized autocorrelation analysis for multi-target detection}
\name{Ye'Ela Shalit$^\star$*, Ran Weber$^\star$*, Asaf Abas$^\dagger$*, Shay Kreymer$^\star$*, and Tamir Bendory$^\star$
\thanks{* These four authors have contributed equally to this work.  \newline
S.K. is supported by the Yitzhak and Chaya Weinstein Research Institute for Signal Processing. T.B. is supported in part by NSF-BSF grant no. 2019752.}}
\address{$^\star$School of Electrical Engineering, Tel Aviv University, Tel Aviv, Israel\\
$^\dagger$Department of Applied Mathematics, Tel Aviv University, Tel Aviv, Israel}

\begin{document}
\ninept

\setlength{\abovedisplayskip}{3pt}
\setlength{\belowdisplayskip}{3pt}

\maketitle
\begin{abstract}
We study the multi-target detection problem of recovering a  target signal from a noisy measurement that contains multiple copies of the signal at unknown locations. Motivated by the structure reconstruction problem in cryo-electron microscopy, we focus on the high noise regime, where noise hampers accurate detection of signal occurrences. Previous works proposed an autocorrelation analysis framework to estimate the signal directly from the measurement, without detecting signal occurrences. Specifically, autocorrelation analysis entails finding a signal that best matches the observable autocorrelations by minimizing a least squares objective. This paper extends this line of research by developing a generalized autocorrelation analysis framework that replaces the least squares by a weighted least squares. The optimal weights can be computed directly from the data and guarantee favorable statistical properties. We demonstrate signal recovery from highly noisy measurements, and show that the proposed framework outperforms autocorrelation analysis in a wide range of parameters.
\end{abstract}
\section{Introduction}
\label{sec:intro}
\begin{figure*}[!tb]
	\begin{subfigure}[ht]{0.33\textwidth}
		\centering
		\includegraphics[width=\columnwidth, height=70pt]{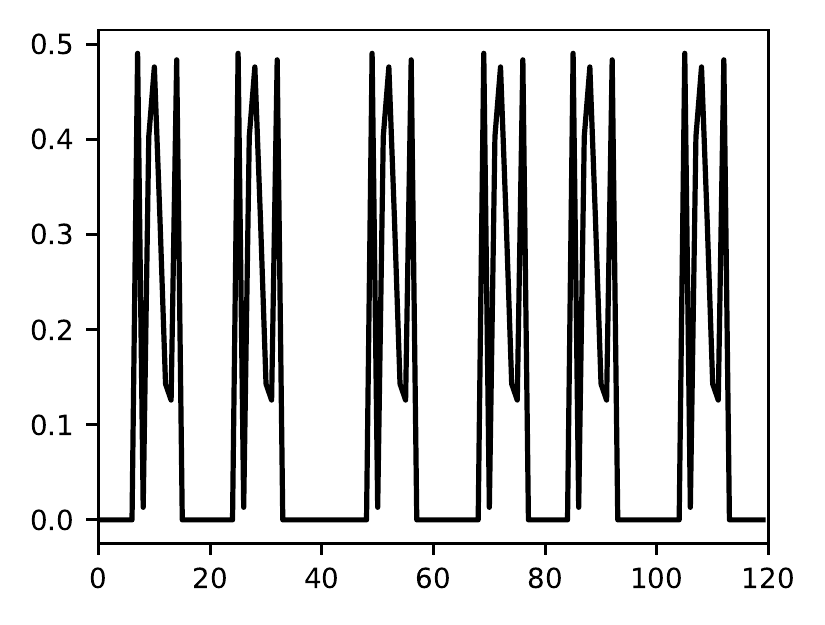}
		\caption{No noise}
	\end{subfigure}
	\hfill
	\begin{subfigure}[ht]{0.33\textwidth}
		\centering
		\includegraphics[width=\columnwidth, height=70pt]{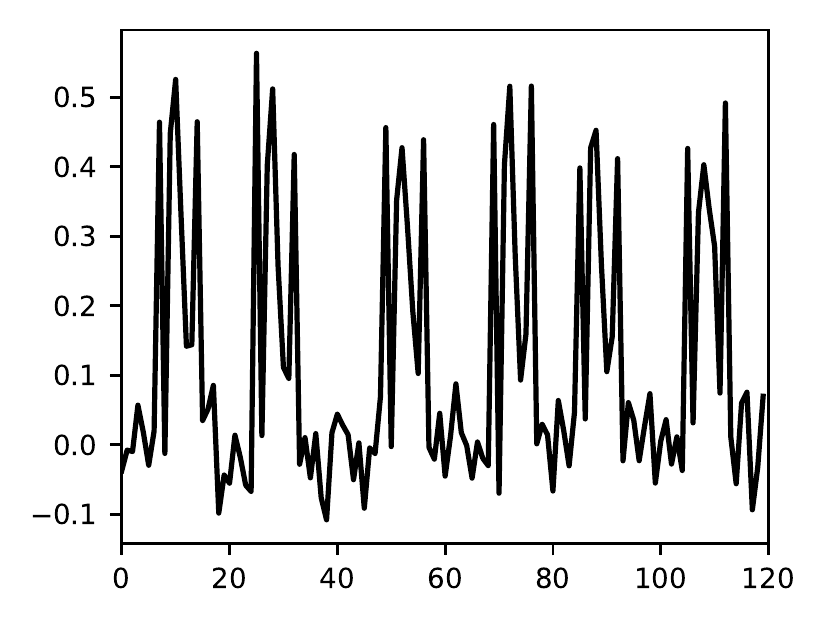}
		\caption{$\text{SNR} = 50$}
	\end{subfigure}
	\hfill
	\begin{subfigure}[ht]{0.33\textwidth}
		\centering
		\includegraphics[width=\columnwidth, height=70pt]{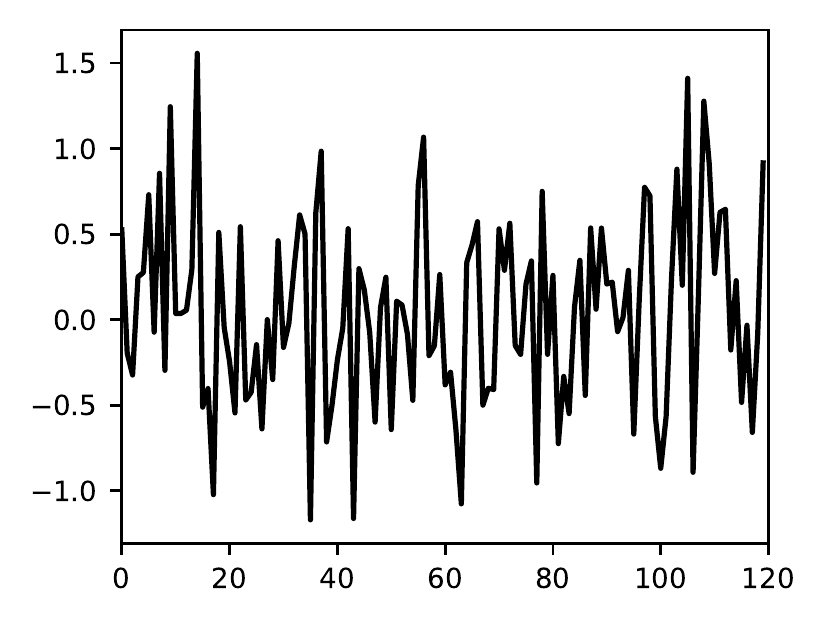}
		\caption{$\text{SNR} = 0.5$}
	\end{subfigure}
	\caption{Three MTD measurements~\eqref{eq:model} {of length~\mbox{$N = 120$}} at different noise levels: (a)~no noise; (b)~\mbox{$\text{SNR} = 50$}; (c)~\mbox{$\text{SNR} = 0.5$}. Each measurement contains six copies of the target signal. Our goal is to estimate the target signal directly from~$y$. We focus on the low SNR regime~(e.g., panel~(c)) in which  signal occurrences are swamped by  noise, and their locations  cannot be reliably detected.}
\label{fig:measurements}
\end{figure*}

We study the multi-target detection (MTD) problem of estimating a target signal~$x \in \mathbb{R}^L$ from a noisy measurement that contains multiple copies of the signal, each randomly translated~\cite{bendory2019multi,lan2020multi,marshall2020image,bendory2021multi,kreymer2021two,bendory2018toward}. Specifically, let~$y \in \mathbb{R}^N$ be a measurement of the form
\begin{equation}
\label{eq:model}
y[\ell] = \sum_{i=1}^{p} x[\ell - \ell_i] + \varepsilon[\ell],
\end{equation}
where \mbox{$\{\ell_i\}_{i=1}^{p} \in \{L + 1, \ldots, N-L\}$} are translations and $\varepsilon[\ell]\overset{\text{i.i.d.}}{\sim} \mathcal{N}(0,\sigma^2)$. The translations and the number of occurrences of~$x$ in~$y$, denoted by~$p$, are unknown. {We assume that the translations are sufficiently separated from each other. Specifically}, each translation is separated by at least a full signal length from its neighbors, {namely},
\begin{equation}
	\label{eq:sep}
	|\ell_{i_1} - \ell_{i_2}| \ge 2L - 1, \quad \text{ for all } i_1 \ne i_2.
\end{equation}
{This model is referred to as the well-separated model~\cite{bendory2019multi}}. Figure~\ref{fig:measurements} presents an example of three  measurements at different signal-to-noise ratios (SNRs). We define~\mbox{$\text{SNR} := \frac{\|x\|_2^2}{L \sigma^2}$}.

The MTD model serves as a mathematical abstraction of the cryo-electron microscopy~(\mbox{cryo-EM}) technology for macromolecular structure determination~\cite{henderson1995potential,nogales2016development,bai2015cryo}. In a \mbox{cryo-EM} experiment \cite{frank2006three}, biological macromolecules suspended in a liquid solution are rapidly frozen into a thin ice layer. An electron beam then passes through the sample, and a two-dimensional tomographic projection is recorded. Importantly, the \mbox{2-D} location and \mbox{3-D} orientation of particles within the ice are random and unknown. This measurement is further affected by high noise levels and the optical configuration of the microscope.

In the current analysis workflow of \mbox{cryo-EM} data \cite{bendory2020single,singer2020computational,scheres2012relion,punjani2017cryosparc}, the~\mbox{2-D} projections are first detected and extracted from the measurement, and later rotationally and translationally aligned to reconstruct the~\mbox{3-D} molecular structure. This approach fails for small molecules, which induce low contrast, and thus low SNR. This makes them difficult to detect and align~\cite{bendory2018toward,singer2020computational,henderson1995potential,aguerrebere2016fundamental}, rendering current \mbox{cryo-EM} algorithmic pipeline ineffective. For example, in the limit~\mbox{$\text{SNR} \rightarrow 0$}, reliable detection of signals' locations within the measurement is impossible~\cite[Proposition~3.1]{bendory2018toward}.

The MTD model was devised in \cite{bendory2018toward} in order to study the recovery of small molecules directly from the measurement, below the current detection limit of \mbox{cryo-EM}~\cite{henderson1995potential,d2021current}. An autocorrelation analysis technique (see Section~\ref{subsec:ac}) was implemented to recover \mbox{low-resolution}~\mbox{3-D} structures from noiseless simulated data under a simplified model. In order to further investigate the method from analytical and computational perspectives, the MTD model was studied in~\cite{bendory2019multi} for one-dimensional signals  under the assumption that the signal occurrences are either {well-separated} or follow a Poisson distribution. In~\cite{lan2020multi}, the mathematical framework was extended to account for arbitrary spacing of signal occurrences. In~\cite{marshall2020image,bendory2021multi}, the authors studied the MTD problem in two dimensions, where the sought images are arbitrarily rotated, but still \mbox{well-separated}, and in~\cite{kreymer2021two} the framework was extended  to an arbitrary spacing distribution of image occurrences.

Autocorrelation analysis is a special case of the method of moments, a classical statistical inference technique, {and was first proposed for cryo-EM data in~\cite{kam1980reconstruction}; see also~\cite{bhamre2015orthogonal,bhamre2017anisotropic,levin20183d}}. It consists of finding a signal that best matches the empirical autocorrelations of the measurement, {usually} by minimizing a least-squares (LS) objective~(see Section~\ref{subsec:ac}). For any noise level, the empirical autocorrelations can be estimated to any desired accuracy for sufficiently large~$N$. Computing the autocorrelations is straightforward and requires only one pass over the data, which is advantageous for massively large datasets, such as \mbox{cryo-EM} datasets~\cite{bendory2020single}. This work studies the application of a \textit{generalized autocorrelation analysis} to the MTD problem. The generalized autocorrelation analysis framework, which is a special case of the generalized method of moments (GMM)~\cite{Hansen1982}, suggests replacing the LS objective of autocorrelation analysis with a weighted LS, and provides a recipe of how to compute the optimal weights; see Section~\ref{subsec:generalized_ac}. The GMM has been proven to be highly effective in a variety of computational tasks, see for example~\cite{wooldridge2001applications,akbar2016more,fan2018optimal,roodman2009xtabond2,abas2021generalized}. We {mention in passing} that the framework can be formulated with other objective functions, rather than LS~\cite{de2002properties}.

The main contribution of this paper is extending the autocorrelation analysis framework introduced in~\cite{bendory2019multi} by developing a generalized autocorrelation analysis framework for the MTD problem. We develop an algorithm for recovering the target signal from a measurement, and demonstrate successful reconstructions in noisy regimes~(see Section~\ref{sec:numerical}). Moreover, we show that the generalized autocorrelation analysis estimator outperforms the classical autocorrelation analysis estimator in {a wide range of} SNRs and measurement lengths. It is thus
a first step towards applying a generalized autocorrelation analysis to recovering small molecules from \mbox{cryo-EM} {experimental datasets}~\cite{bendory2018toward}.

\section{Computational framework}
\label{sec:math}
\subsection{Autocorrelation analysis}
\label{subsec:ac}
{Before introducing the generalized autocorrelation analysis framework, we begin by presenting a classical autocorrelation analysis}. The autocorrelation of order~$q$ of a signal~\mbox{$z \in \mathbb{R}^{N}$} is defined as
\begin{equation*}
A_z^q[\ell_1, \ldots, \ell_{q-1}] := \mathbb{E}_z\Big[\frac{1}{N} \sum_{i \in \mathbb{Z}} z[i] z[i + \ell_1] \cdots z[i + \ell_{q-1}]\Big],
\end{equation*}
where~$\ell_1, \ldots, \ell_{q-1}$ are integer shifts. Indexing out of bounds is zero-padded, that is,~\mbox{$z[i] = 0$} out of the range~\mbox{$\{0, \ldots, {N-1}\}$}. As~$N$ grows indefinitely, by the law of large numbers, the empirical autocorrelations of~$z$ almost surely (a.s.) converge to the population autocorrelations of~$z$:
\begin{equation}\label{eq:genericAC}
\lim_{N \rightarrow \infty} \frac{1}{N} \sum_{i \in \mathbb{Z}} z[i] z[i + \ell_1] \cdots z[i + \ell_{q-1}] \stackrel{\text{a.s.}}{=}A_z^q[\ell_1, \ldots, \ell_{q-1}].
\end{equation}

To formulate an autocorrelation analysis, we write the first three autocorrelations of~$y$. {We use the first three autocorrelations since the third-order autocorrelation is the lowest-order autocorrelation that determines a generic signal uniquely~\cite{bendory2019multi}}. According to~\eqref{eq:genericAC}, the first-order autocorrelation is defined as
\begin{equation*}
A_{y}^1 := \frac{1}{N} \sum_{i \in \mathbb{Z}} y[i],
\end{equation*}
which is just the mean of the measurement. The second-order autocorrelation of~$y$, \mbox{$A_{y}^2: \mathbb{Z} \rightarrow \mathbb{R}$}, is defined by
\begin{equation*}
A_{y}^2 [\ell_1] := \frac{1}{N} \sum_{i \in \mathbb{Z}} y[i] y[i + \ell_1].
\end{equation*}
{For almost all \mbox{1-D} signals, the second-order autocorrelation does not contain enough information to determine a signal uniquely~\cite{beinert2018enforcing,bendory2017fourier}}. The third-order autocorrelation~\mbox{$A_{y}^3: \mathbb{Z} \times \mathbb{Z} \rightarrow \mathbb{R}$} {is given} by
\begin{equation*}
A_{y}^3 [\ell_1, \ell_2] := \frac{1}{N} \sum_{i \in \mathbb{Z}} y[i] y[i + \ell_1] y[i + \ell_2].
\end{equation*}

Next, we want to relate the observable autocorrelations $A_y^1$, $A_y^2$, and~$A_y^3$, to the signal~$x$, under the statistical model~\eqref{eq:model}.
Importantly, under the well-separated condition~\eqref{eq:sep}, for translations in the range \mbox{$\mathcal{L} = \{0, \ldots, {L - 1}\}$}, any given occurrence of~$x$ in~$y$ is only ever correlated with itself, and never with another occurrence.
In~\cite{bendory2019multi}, it was shown that under the well-separated condition~(\ref{eq:sep}), for any fixed level of noise~$\sigma^2$, density~$\gamma$ and signal length~$L$, in the limit~\mbox{$N \rightarrow \infty$}, the autocorrelations of the measurement are equal,  up to some {predicted} bias terms, to the autocorrelations of the signal times a density constant. Specifically, we have that
\begin{align}
\label{eq:well_separated_1st}
A_{y}^1 &\stackrel{\text{a.s.}}{=} \gamma A_{x}^1, \\
\label{eq:well_separated_2nd}
A_{y}^2 [\ell_1] &\stackrel{\text{a.s.}}{=} \gamma A_{x}^2 [\ell_1] + \sigma^2\delta[\ell_1], \\
\label{eq:well_separated_3rd}
A_{y}^3 [\ell_1, \ell_2] &\stackrel{\text{a.s.}}{=} \gamma A_{x}^3 [\ell_1, \ell_2]  + \gamma A_{x}^1 B[\ell_1, \ell_2],
\end{align}
for~$\ell_1, \ell_2 \in \mathcal{L}$, {where~\mbox{$B[\ell_1, \ell_2] := \sigma^2 (\delta[\ell_1] + \delta[\ell_2] + \delta[\ell_1 - \ell_2])$}}, and $\delta$
is the Kronecker delta function defined by $\delta[0]=1$ and $\delta[\ell\neq 0]=0$.
Here,~$\gamma$ is the density of the target images in the measurement and is defined by
\begin{equation*}
\gamma = p \frac{L}{N}.
\end{equation*}
For instance, the measurements in Figure~\ref{fig:measurements} correspond to~\mbox{$\gamma = 0.4$}. {We mention that if the signal occurrences violate the separation condition~(\ref{eq:sep}) but follow a Poisson distribution, the autocorrelations are equivalent to~(\ref{eq:well_separated_1st}) -~(\ref{eq:well_separated_3rd})~\cite{bendory2019multi}}.

Notably, the relations between the autocorrelations of~$y$ and~$x$ do not directly depend on the location of individual signal occurrences in the measurement, but only through the density parameter~$\gamma$. Therefore, detecting the signal occurrences is not a prerequisite for signal recovery, and thus signal recovery is possible even in very low SNR regimes.

In \cite{bendory2019multi,lan2020multi,marshall2020image,bendory2021multi,kreymer2021two}, it was suggested to find the signal that best matches the observable autocorrelations by minimizing a LS objective. Specifically, using~\eqref{eq:well_separated_1st} - \eqref{eq:well_separated_3rd} the LS reads
\begin{align}
\label{eq:optimization}
\argmin_{x\in\mathbb{R}^L, \gamma > 0} & (A_y^1 - \gamma A_x^1)^2  + w_2 \sum_{\ell_1 = 0}^{L - 1} \|A_y^2[\ell_1] - \gamma A_x^2[\ell_1] - \sigma^2 \delta[\ell_1]\|_2^2\nonumber\\ +& w_3 \sum_{\ell_1 = 0}^{L - 1} \sum_{\ell_2 = 0}^{L - 1} \|A_y^3[\ell_1, \ell_2] - \gamma A_x^3[\ell_1, \ell_2] - \gamma A_{x}^1 B[\ell_1, \ell_2]\|_2^2,
\end{align}
where the weights~$w_2$ and~$w_3$ were chosen such that each term is equally weighted, as suggested by~\cite{bendory2019multi}, and $B[\ell_1, \ell_2]$ is defined in~\eqref{eq:well_separated_3rd}. We note that the optimization problem is non-convex, and thus there is no guarantee to converge to a global optimum. Nevertheless, similarly to previous papers on MTD, our numerical results (see Section~\ref{sec:numerical}) suggest that standard gradient-based methods succeed in recovering~$x$ from only a few random initial guesses.

\subsection{Generalized autocorrelation analysis}
\label{subsec:generalized_ac}
The generalized autocorrelation analysis framework is a special case of the GMM.  In its most simplified form, the GMM generalizes the method of moments by replacing the LS objective function by  weighted LS with  optimal weights. A specific  choice of weights guarantees favorable asymptotic statistical properties, such as  minimal asymptotic variance of the estimation error~\cite{Hansen1982} {(see Section~\ref{gmm:large})}. In this work, we adapt the GMM to autocorrelation analysis and term the method generalized autocorrelation analysis.

Let us define the moment function~$f(\theta, y)$  for $\theta\in\Theta$, where $\Theta$ is a compact parameter space.
 The moment function is chosen such that its expectation is zero only at a single point~$\theta=\theta_0$, where~$\theta_0$ is the ground truth parameter (in our case, the target signal~$x$ and the density parameter~$\gamma$). Namely,
\begin{equation}\label{Eq-GMM-1}
	\E\left[f(\theta,y)\right] = 0 \quad \text{if and only if} \quad \theta = \theta_0.
\end{equation}
Since the moment function need only to satisfy the uniqueness condition~\eqref{Eq-GMM-1} and a few additional mild regularity conditions (which can be found in~\cite{Hansen1982,abas2021generalized,Hall2005}), the GMM was applied to a wide range of estimation problems,  such as panel data problem~\cite{blundell2000gmm} and subspace estimation~\cite{fan2018optimal}.

In order to define the moment function for the MTD problem, we first define the $i$-th observation from the measurement~$y$ as follows:
\begin{equation}\label{eq:sample}
	y_i := [y[i],\ldots, y[i+L-1]]\in\R^L.
\end{equation}
A natural choice {of a} moment function~$f(\theta, y_i)$, for~$y_i$ from~\eqref{eq:sample}, {is the discrepancy between the autocorrelations of~$y_i$ and the population autocorrelations}:
\begin{multline} \label{Eq-GMM-2}
	f(\theta, y_i) := \\
	\begin{bmatrix}
		&\gamma A_x^1 - A_{y_i}^1\\
		&\left\{\gamma A_x^2[\ell_1] + \sigma^2 \delta[\ell_1] - A_{y_i}^2 [\ell_1]\right\}_{\ell_1 = 0}^{L-1} \\
		&\left\{\gamma A_x^3[\ell_1, \ell_2] + \gamma A_{x}^1 B(x, \ell_1, \ell_2) - A_{y_i}^3[\ell_1, \ell_2]\right\}_{\ell_1, \ell_2 = 0}^{L-1}
	\end{bmatrix},
\end{multline}
where~\mbox{$\theta := [x, \gamma]$}. For convenience, we treat each autocorrelation as a column vector
and the right hand side of~\eqref{Eq-GMM-2} as their concatenation.
Notice that the chosen moment function~\eqref{Eq-GMM-2} fulfills the {uniqueness} condition~\eqref{Eq-GMM-1} for generic signals~\cite{bendory2019multi}. The estimated sample moment function is the average of~$f$ over~$N$ observations:
\begin{equation}\label{Eq-2-5}
	g_N(\theta) = \frac{1}{N} \sum_{i = 0}^{N - 1} f(\theta, y_i).
\end{equation}
The generalized autocorrelation estimator is defined as the minimizer of the weighted LS objective
\begin{equation} \label{eq:opt_theta}
	\hat{\theta}_N = \arg\min_{\theta \in \Theta} \ g_N(\theta)^T W_N g_N(\theta).
\end{equation}
Here, $W_N$ is a fixed positive semi-definite (PSD) matrix. Note that the LS estimator~(\ref{eq:optimization}) is a special case of~(\ref{eq:opt_theta}), where~$W_N$ is a diagonal matrix with the weights from~\eqref{eq:optimization}.
\begin{figure*}[!tb]
	\begin{subfigure}[ht]{0.245\textwidth}
		\centering
		\includegraphics[width=\columnwidth]{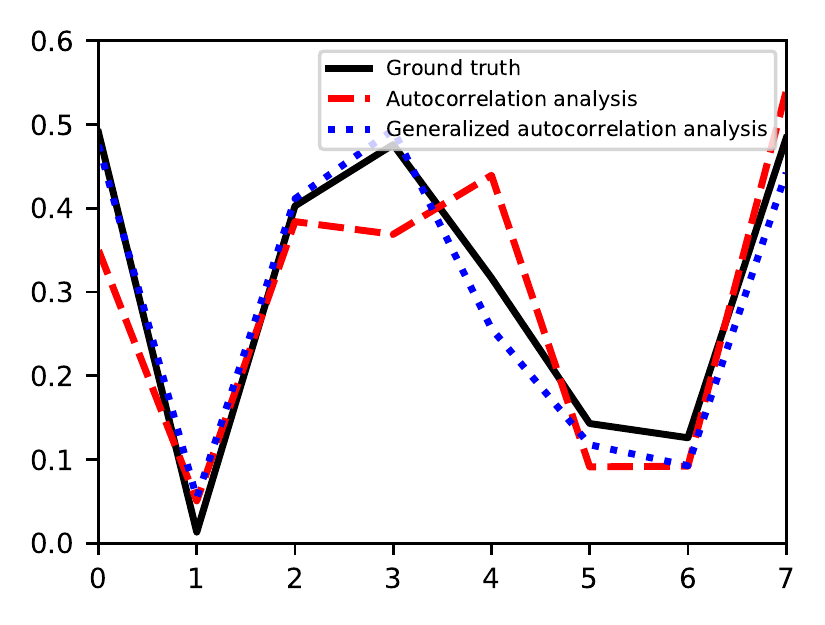}
		\caption{$N = 10^5$}
	\end{subfigure}
	\hfill
	\begin{subfigure}[ht]{0.245\textwidth}
		\centering
		\includegraphics[width=\columnwidth]{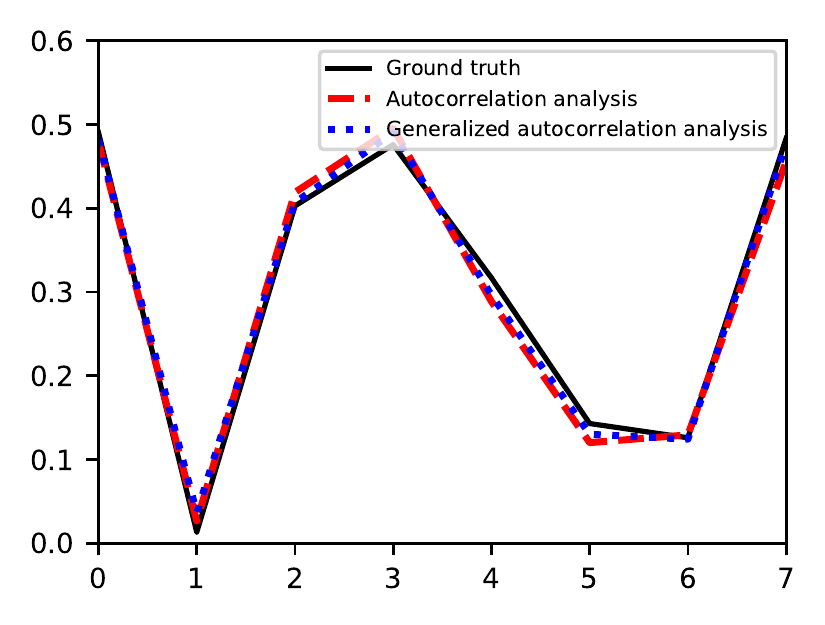}
		\caption{$N = 10^6$}
	\end{subfigure}
	\hfill
	\begin{subfigure}[ht]{0.245\textwidth}
		\centering
		\includegraphics[width=\columnwidth]{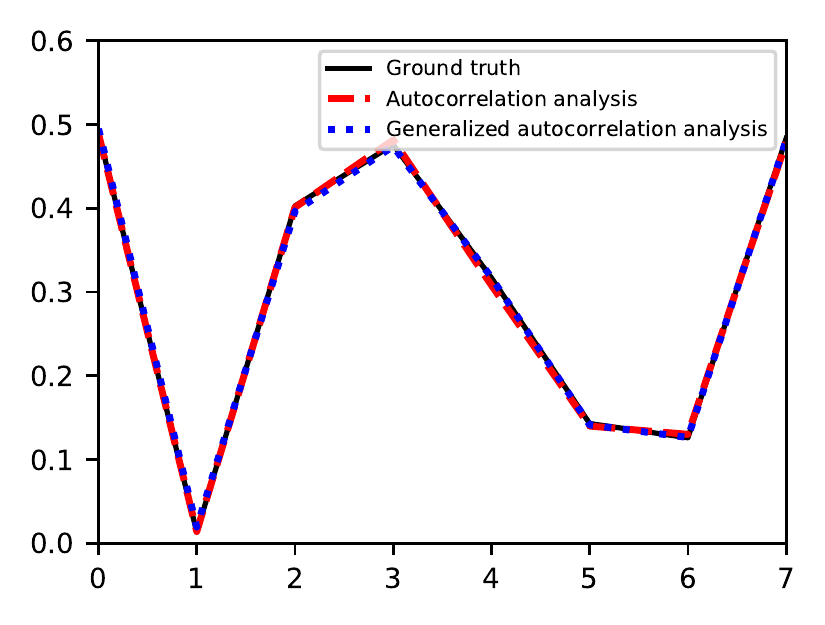}
		\caption{$N = 10^7$}
	\end{subfigure}
	\hfill
	\begin{subfigure}[ht]{0.245\textwidth}
		\centering
		\includegraphics[width=\columnwidth]{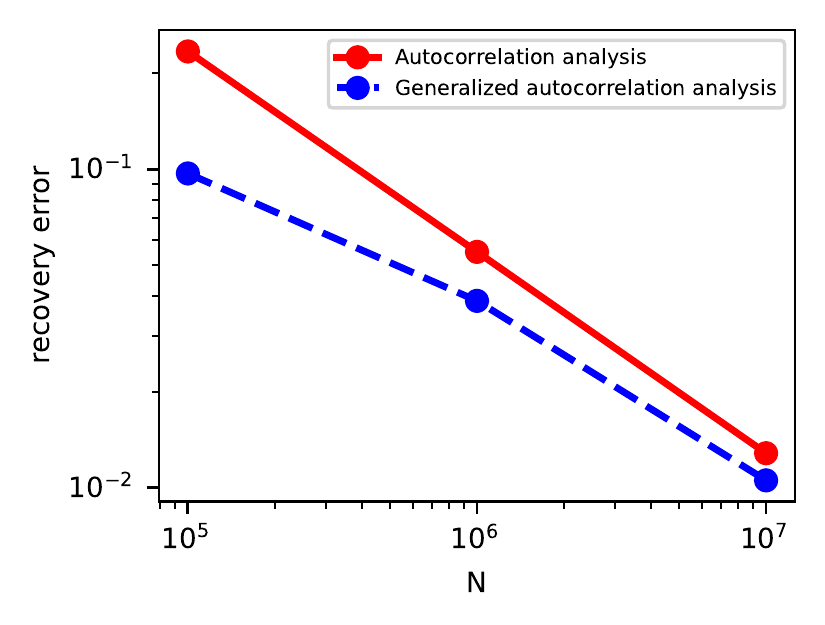}
		\caption{Recovery error}
	\end{subfigure}
	\caption{Panels (a), (b), and (c) present signal estimates using autocorrelation analysis (in red) and generalized autocorrelation analysis (in blue) with different measurement lengths. Panel (d) presents the recovery error using both methods as a function of measurement length~{($N$)}.}
\label{fig:recoveries}
\end{figure*}

\subsection{Large sample properties}
\label{gmm:large}
Before presenting the statistical properties of the generalized autocorrelation analysis estimator, we fix notation. We denote by~$\overset{p}{\to}$ and~$\overset{d}{\to}$ convergence in  probability and in distribution, respectively. Let
\begin{equation} \label{eqn:cov_mat_S}
	S := \lim_{N\to \infty}\Cov\left[\sqrt{N}g_N(\theta_0)\right],
\end{equation}
be the covariance matrix of the estimated sample moment function~(\ref{Eq-2-5}) at the ground truth~$\theta_0$. We denote by $\{W_N\}_{N=1}^\infty$ a sequence of PSD matrices which converges almost surely to a positive definite matrix~$W$. The expectation of the Jacobian of the moment function at the ground truth~$\theta_0$ is denoted by $G_0 = \E\left[\partial f(\theta_0, y) / \partial \theta^T\right]$.

The large sample properties of the GMM estimator, and thus also of the generalized autocorrelation estimator,  were derived in~\cite{Hansen1982}, and are presented in the following theorem.

\begin{theorem}\label{Thm-2-6}
	{Under condition~(\ref{Eq-GMM-1}) and a few additional mild regularity conditions that can be found in~\cite{Hansen1982,abas2021generalized,Hall2005}}, the GMM estimator (and thus also the generalized autocorrelation analysis estimator) satisfies:
	\begin{enumerate}[label={\Alph*}.]
		\item  \label{Thm-2-2}
		\textnormal{(Consistency)} $\hat{\theta}_N \overset{p}{\to} \theta_0$.

		\item \label{Thm-2-3} \textnormal{(Asymptotic normality)}
		\[\sqrt{N} ( \hat{\theta}_N - \theta_0) \overset{d}{\to} \mathcal{N}(0, M S M^T ),\] where $M =[G_0^T W  G_0]^{-1} G_0^T  W$.

		\item \label{Thm-2-5} \textnormal{(Optimal choice of a weighting matrix)} The minimum asymptotic variance of $\hat{\theta}_N$ is given by $(G_0^T S^{-1} G_0)^{-1}$ and is attained by $W = S^{-1}$.
	\end{enumerate}
\end{theorem}

Theorem~\ref{Thm-2-6} {shows that the matrix~$W = S^{-1}$ guarantees} a minimal asymptotic variance of the estimator’s error. The covariance matrix~$S$ {of~(\ref{eqn:cov_mat_S}), which plays a central role in Theorem~\ref{Thm-2-6},} is required to be a positive definite matrix. Therefore, the moment function must be chosen so that~$S$ is full-rank. Empirically, in our case the matrix~$S$ is indeed full-rank when removing repeating entries of $f$ that appear due to  inherent symmetries of the autocorrelations~\cite{abas2021generalized}. We note that in general the matrix~$W$ depends on the ground truth~$\theta_0$, and thus cannot be computed from the data. {Therefore, it is common to use iterative methods that alternate between optimizing~(\ref{eq:opt_theta}) given~$W$, and computing~$W$ given the current estimate of~$\theta$}. However, for our specific choice of moment function~(\ref{Eq-GMM-2}), the matrix~$S$ is independent of the parameters of interest~\cite{abas2021generalized}, and thus can be computed directly from the data:
\begin{multline*}
	\Cov[g(\theta)] = \\ \Cov\left[\left\{[A_{y_i}^1;\left\{A_{y_i}^2[\ell_1]\right\}_{\ell_1=0}^{L-1};\left\{A_{y_i}^3[\ell_1, \ell_2]\right\}_{\ell_1, \ell_2 = 0}^{L-1}]\right\}_{i = 0}^{N - 1}\right].
\end{multline*}

\section{Numerical experiments}
\label{sec:numerical}
This section compares the numerical performance of the generalized autocorrelation analysis framework and the classical autocorrelation analysis method. A comparison of autocorrelation analysis {to a naive method that detects and extracts the signal occurrences, and then averages, was} conducted in~\cite{lan2020multi,kreymer2021two}. While this method works well in a high SNR environment, it fails in low SNR regimes---the focal point of this paper---and thus omitted from this section. The optimization problem~(\ref{eq:optimization}) was minimized using the~\mbox{Broyden-Fletcher-Goldfarb-Shanno}~(BFGS) algorithm, while ignoring the positivity constraint on~$\gamma$ (namely, treating it as an unconstrained problem).  We measure the estimation error by
\begin{equation*}
\text{error}(x) = \frac{\|x - x^*\|_2}{\|x^*\|_2},
\end{equation*}
where~$x^*$ is the true signal, and~$x$ is the estimated signal. In the experiments presented in Sections~\ref{subsec:exp_size} and~\ref{subsec:exp_SNR}, the measurements were generated according to~(\ref{eq:model}) with density~\mbox{$\gamma = 0.2$}, and the target signals are of length~\mbox{$L = 21$}. Each entry of the signals was drawn i.i.d.\ from a uniform distribution on~$[0,1]$, and the target signals were normalized such that~$\|x\|_2  =1$. We estimated the signal from~$5$ random initial guesses (that were drawn from the same distribution as the ground truth signal) and~\mbox{$\gamma_{\text{init}} = 0.18$}, and calculated the estimation error of the {signal} estimate whose final objective function is minimal. Figures~\ref{fig:err_size_experiment} and~\ref{fig:err_noise_experiment} present the median error over~$50$ trials. The code to reproduce all experiments is publicly available at~\url{https://github.com/krshay/MTD-GMM}.

\subsection{Recovery from a noisy measurement}
\label{subsec:exp_recovery}
In Figure~\ref{fig:recoveries} we present a successful recovery of a target signal of length~\mbox{$L = 8$} from a noisy measurement with~\mbox{$\text{SNR} = 0.5$} and~\mbox{$\gamma = 0.2$}, using autocorrelation analysis and generalized autocorrelation analysis. The noise level is visualized in Figure~\ref{fig:measurements}.
As expected, the recovery error decreases  as the measurement length increases. Remarkably, for all measurement lengths~$N$, the generalized autocorrelation analysis framework shows superior numerical performance.

\subsection{Recovery error as a function of the measurement length}
\label{subsec:exp_size}
\begin{figure}[!tb]
	\begin{subfigure}[ht]{\columnwidth}
		\centering
		\includegraphics[width=0.8\columnwidth, keepaspectratio]{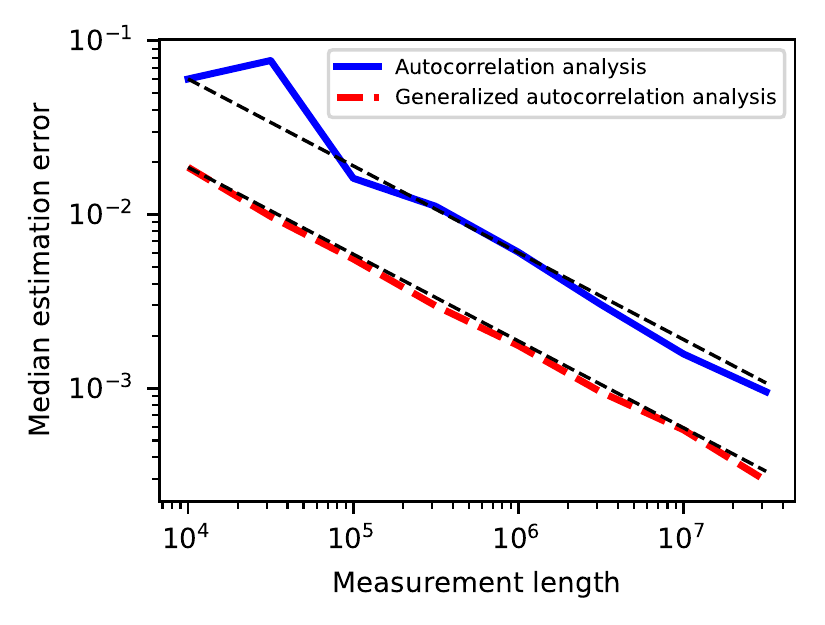}
	\end{subfigure}
	\caption{The median estimation error as a function of the measurement length~$N$ with $\text{SNR}=50$, by: (a) autocorrelation analysis; (b)~generalized autocorrelation analysis. {The black dashed lines illustrates a slope of~$-1/2$, as predicted by the law of large numbers.}}
	\label{fig:err_size_experiment}
\end{figure}

Figure~\ref{fig:err_size_experiment} presents recovery error as a function of the measurement length~$N$. We set~\mbox{$\text{SNR} = 50$}, demonstrating a high SNR environment. As expected by the law of large numbers, the recovery error of both estimators decays as~$N^{-1/2}$. Evidently, the generalized autocorrelation analysis estimator significantly outperforms {the} classical autocorrelation analysis estimator for all~$N$.

\subsection{Recovery error as a function of the SNR}
\label{subsec:exp_SNR}
\begin{figure}[!tb]
	\begin{subfigure}[ht]{\columnwidth}
		\centering
		\includegraphics[width=0.8\columnwidth, keepaspectratio]{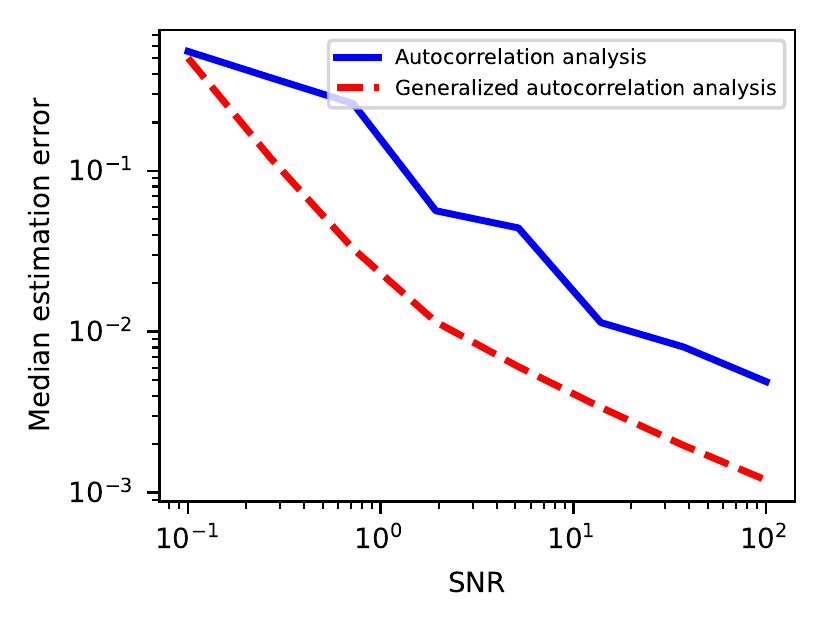}
	\end{subfigure}
	\caption{The median estimation error as a function of SNR, for measurements with length~\mbox{$N = 10^6$}, by: (a)~autocorrelation analysis; (b)~generalized autocorrelation analysis estimator. Evidently, the generalized autocorrelation analysis estimator outperforms classical autocorrelation analysis for all SNR levels.}
	\label{fig:err_noise_experiment}
\end{figure}

Figure~\ref{fig:err_noise_experiment} presents recovery error as a function of {the} SNR, for a fixed measurement length~$N = 10^6$. For all levels of SNR, the recovery error using  generalized autocorrelation analysis  is smaller than the error of standard autocorrelation analysis. In addition, the slope of the error curve increases dramatically at low SNR, around~\mbox{$\text{SNR} \approx 1$}, which is a known phenomenon in the \mbox{cryo-EM} literature, see for example~\cite{sigworth1998maximum,abbe2018multireference,perry2019sample}.

\section{Conclusion}
\label{sec:conclusion}
This paper is motivated by the effort of reconstructing small~\mbox{3-D} molecular structures using \mbox{cryo-EM}, below the current detection limit. The main contribution of this study is incorporating the generalized method of moments into the computational framework proposed in~\cite{bendory2018toward} for the MTD problem. Theorem~\ref{Thm-2-6} shows the optimality of the proposed framework, and it is corroborated by numerical experiments.

Future work includes extending the generalized autocorrelation analysis estimator to the~\mbox{2-D} and~\mbox{3-D} cases of the MTD problem~\cite{bendory2018toward}. Furthermore, we wish to generalize the framework to the case of arbitrary spacing distribution between signal occurrences (namely, signals that violate the separation condition~\eqref{eq:sep})~\cite{lan2020multi,kreymer2021two}, with the {ultimate} goal of applying the method to \mbox{cryo-EM} {experimental} datasets.

\vfill
\newpage

\end{document}